\documentclass[twocolumn,
aps,nofootinbib,showpacs,showkeys,preprint
tightenlines
] {revtex4}

\usepackage{epsf,epsfig,subfigure,graphicx,amsmath,amssymb}
\usepackage{color}
\newcommand{\dis}[1]{\begin{equation}\begin{split}#1\end{split}\end{equation}}

\newcommand{\dslash}{{\partial\hskip -0.20cm\slash}}
\newcommand{\kslash}{{k\hskip -0.20cm\slash}}

\newcommand{\tev}{\,\textrm{TeV}}
\newcommand{\gev}{\,\textrm{GeV}}
\newcommand{\NDMMSSM}{$N_{\textrm{DM}}$MSSM}
\newcommand{\MG}{{$M_{\rm GUT}$}}
\newcommand{\ie}{{\it i.e.}\ }
\newcommand{\etal}{{\it et al.}\ }

\begin{document}

\begin{flushright}{\tt ~~\\~~\\ SNUTP 09-010 \quad\quad\quad}\end{flushright}

\title{\Large\bf Decaying dark matter with heavy axino }

\author{   Ji-Haeng Huh and
Jihn E. Kim\email{jekim@ctp.snu.ac.kr}%  and Bumseok Kyae
}
\affiliation{ Department of Physics and Astronomy and Center for Theoretical Physics, Seoul National University, Seoul 151-747, Korea
 }

\begin{abstract}
A TeV scale decaying dark matter chiral multiplet $N$ is introduced in addition to the minimal supersymmetric standard model(MSSM). For a {\it calculable} abundance of $N$, we introduce heavy axino decaying to $N$ and MSSM particles including the lightest supersymmetric particle(LSP). In the scenario that heavy axino, once dominating the energy density of the universe, decays after the LSP decouples, it is possible to estimate the relative cosmic abundances of $N$ and  the LSP. Dimension 6 interactions allow the lifetime of the fermionic or the bosonic superpartner of $N$ in the $10^{27}$ s range to be compatible with the recent astrophysical bounds. A diagrammatic strategy obtaining a suppression factor $1/M^2$ is also given.
\end{abstract}


\maketitle

%%%%%%%%%%%%%%%%%%%%%%%%%%%%%%%%%%%%%%%%%%%%%%%%%%%%%%%%%%%%
%%%%%%%%%%%%%%%%%%%%%%%%%%%%%%%%%%%%%%%%%%%%%%%%%%%%%%%%%%%%
%{\it Introduction~}---
\section{Introduction}
Recent observations of high energy galactic positrons, electrons, antiprotons, and gamma rays attracted a great deal of attention on dark matter(DM) scenarios. If TeV scale decaying DM(DDM) decays at the present epoch to produce these high energy particles, it needs a very long lifetime (of order $10^{27}$ s) so that its decay is within the allowed limits of experimental observations. On the theoretical side, the standard model(SM) has been extended to the MSSM, mainly to solve the gauge hierarchy problem. This supersymmetric(SUSY) extension has found another bonuses: the existence of the LSP $\chi$ as a cold DM candidate and the gauge coupling unification around $(2-3)\times 10^{16}$ GeV. In Ref. \cite{HuhKK08}, a further extension of the MSSM by an additional DM component $N$ (called \NDMMSSM) to produce enough high energy positrons was suggested together with charged SM singlets $E^\pm$ to explain  PAMELA's excess positrons \cite{PAMELAe} from $N+\chi$ annihilation. Interestingly, grand unified theories(GUTs) allowing charged SM singlets $E^\pm$ are possible in the flipped SU(5) GUT \cite{FlippedSU5}, which has an ultraviolet completion in the heterotic string \cite{flipstring}.

Later last year, the ATIC data raised the DM scale up to TeV \cite{ATICnature}, and the genie for TeV scale DDM has been let free. With TeV DDM $N$, the charged singlet $E^\pm$ of Ref. \cite{HuhKK08} may or may not be needed below the mass scale $m_N$ but \NDMMSSM~ can still be considered. However, the recent {\it Fermi} LAT data is in conflict with the ATIC data of several hundred GeV electrons \cite{Fermi1}. Even though the TeV scale cosmic-ray(CR) electrons are explained by the known astrophysical backgrounds, PAMELA's CR positron excess at the 10--80 GeV range may need another contribution beyond the known backgrounds \cite{Profumo09}. On the other hand, PAMELA's low antiproton flux \cite{PAMELAp} has been generally regarded as a difficulty of DDM scenario \cite{Gondolo08,HuhKK08}. Note however that the old background estimates of the H.E.S.S. data \cite{HESS08} had large systematic uncertainties. For example, Ref. \cite{Bergstrom09} considered these uncertainties to allow a leptonic background smaller by a factor 0.85 of the old background value. If one applies this argument to antiproton flux also, one can allow some antiprotons from DDM decay. Interestingly, PAMELA's CR antiproton flux above 10 GeV has the same shape as the old background estimate, which may be interpreted as ``the old estimate in fact contained extra antiprotons". This new explanation of the old antiproton background allows a room for antiproton injection to the galaxic DM soup from DDM decay. So, models producing some antiprotons in addition to positrons need not be ruled out from the outset.

With this new perspective, now it is very interesting to consider the TeV scale DDM possibility, even allowing some antiproton flux from the DDM decay though we will skip the discussion of the antiproton flux in this paper.
In this spirit, we consider the TeV scale DDM possibility by the simplest extension of the MSSM with just one chiral multiplet $N$ at the next mass level beyond the MSSM, which is an \NDMMSSM~ model \cite{HuhKK08}. The supermultiplet $N$ contains the bosonic partner $\tilde N$ and the fermionic partner $N$.\footnote{Without confusion, we use the same notation $N$ for the supermultiplet and its fermionic partner.} The chiral field $N$ becomes a two-component massive Majorana fermion at the true vacuum.  The LSP $\chi$ is assumed to be stable with the unbroken R-parity and may constitute a dominant portion of galactic DM. Then, the TeV scale DDM $N$ can decay to MSSM particles. The needed range of the $N$ lifetime with the stable LSP $\chi$ is $\sim(m_N/m_\chi)10^{26}$ s.\footnote{ The lifetime as a function of $m_N/m_\chi$ for two DM components can be gleaned from \cite{Kyae09}.} The number density of the $N$ chiral multiplet is completely unknown at this point. But, if some heavier particle $\tilde X$ dominates the energy density of the universe and decays to both $N$ and $\chi$ below the LSP decoupling temperature, it is possible to estimate the relative abundances of $N$ and $\chi$. We explore this possibility, interpreting $\tilde X$ as the axino \cite{ChoiKY08}.

The axion has the anomalous coupling to gluons. So, the heavy axino enables us to estimate the relative abundances of $N$ and $\chi$ through the anomalous coupling and a superpotential term,
\begin{equation}
\int d^2\vartheta \left(\frac{1}{4M'}NNXX-\frac{c_g\alpha_g}{4\sqrt2\pi } \vartheta_g{\cal W}_{g}{\cal W}_g
\right)\label{eq:axinodecay}
\end{equation}
where $c_g$ are coefficients of O(1), $\alpha_g$ are the gauge couplings, and $\vartheta_g$s are the vacuum angle terms. $\vartheta_3$ defines the axion: $c_3\vartheta_3= X/F_a$ \cite{KimCarosi}. The relevant axino decay
Lagrangian \cite{eq:axinocoupl} is $(\langle X\rangle/M')N\tilde N\tilde a+(\alpha_3/4\sqrt2\pi F_a)\tilde G \sigma^{\mu\nu}G_{\mu\nu}\tilde a$ where $\tilde G$ is  the gluino and $G^{\mu\nu}$ is the gluon field strength. Here, we neglect the coupling $\phi_u\phi_d XX/4M'$, assuming that the LSP is predominantly bino. One gluino will produce one LSP in the end, and hence we expect the following $N$ and $\chi$ ratio from the axino decay, in the limit $m_N\gg m_\chi$,
\begin{eqnarray}
&&\frac{{\rm Number~of~}N}{{\rm Number~of~}\chi}\simeq 2\left(\frac{32\pi^2}{\alpha_3^2}\right)\left(\frac{\langle X\rangle}{M'}\right)^2.\label{eq:Nchiratio}
\end{eqnarray}
To obtain this ratio at the level of $\sim m_\chi/m_N\sim 10^{-2}$ and  $F_a\sim 4\times 10^{11}$ GeV, we need $M'\sim 2\times 10^{15}$ GeV which falls in a broad GUT scale with our notation of \MG$\sim  10^{15}-5\times 10^{16}$ GeV.

%%%%%%%%%%%%%%%%%%%%%%%%%%%%%%%%%%%%%%%%%%%%%%%%%%%%%%%%%%%%
%%%%%%%%%%%%%%%%%%%%%%%%%%%%%%%%%%%%%%%%%%%%%%%%%%%%%%%%%%%%
%{\it Models~}---
\section{Models}
In addition to the MSSM symmetries we introduce the R-parity and the Peccei-Quinn(PQ) symmetry U(1)$_\Gamma$. In addition, we also introduce {\it  matter parity $P$}. The attractive feature of the PQ symmetry is that it solves the strong CP problem, the resulting invisible axion may constitute a cold DM component, and its breaking scale is narrowed down to a window\footnote{But, note that there exists the possibility that $F_a$ can be larger than $10^{12}$ GeV for a small initial misalignment angle \cite{BaeFa98}. } $10^9\le F_a\le 10^{12}$ GeV \cite{KimCarosi} so that our estimate of the $N$ lifetime is more or less predictive.

\begin{figure}[!]
\begin{center}
\resizebox{0.5\columnwidth}{!}
{\includegraphics{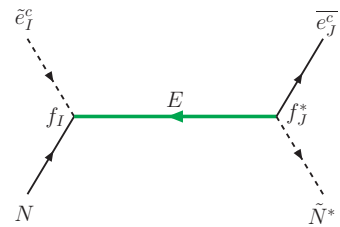}}\vskip 0.1cm
\centerline{(a)}\vskip 0.3cm
\resizebox{0.5\columnwidth}{!}
{\includegraphics{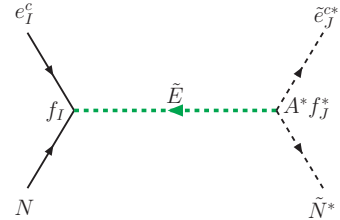}}\vskip 0.1cm
\centerline{(b)}
\end{center}
\caption{Possible operators for $N$ decay suppressed with a super-heavy mass $1/M^2$: (a) The propagator with one arrow accompanies a momentum in the numerators of the propagator, and (b) A diagram with the $A$-term. }\label{fig:dim6HKK}
\end{figure}

The simplest $1/M^2$ suppression results with four external fields which however cannot be expressed as a superpotential term. This interaction includes a derivative coupling. An example of the derivative interaction with four external lines is given in Fig. \ref{fig:dim6HKK}(a) with the coupling $e_I^cE N$ of Ref. \cite{HuhKK08}. The Weyl field propagator for one direction arrow is $i\kslash_E/(k_E^2-m_E^2)$ and Fig.  \ref{fig:dim6HKK}(a) gives a dimension 6 operator with one derivative multiplied by $1/m_E^2$. In contrast, two colliding Weyl fields gives a Majorana mass $m_n$ in the numerator of a propagator as $im_n/(k^2-m_n^2)$.  For Fig. \ref{fig:dim6HKK}(a), we have an operator for $N$ decay, \ie the interaction Lagrangian becomes
\begin{equation}
f_If_J^* \overline{e_J^c}
\frac{i\kslash_{E}}{k^2_{E}-m^2_E}
N  \tilde N^* \tilde e_I^c\to
\frac{ \tilde N^*}{M^2} (\overline{e_J^c}\dslash)
N  \tilde e_I^c  \label{eq:effderInt}
\end{equation}
where the super-heavy mass is $M^2=m^2_E/f_If_J^*$.
By the interaction (\ref{eq:effderInt}), $\tilde N$ (or $N$) decays to $N$ (or $\tilde N$) if $\tilde N$ ($N$) is heavier than $N$ ($\tilde N$).
The dimension 6 interaction can also arise with five external lines expressible as $
({f_{\alpha\beta\gamma}}/{M^2}) \phi_{\alpha,\rm boson} \phi_{\beta,\rm boson}\phi_{\gamma,\rm fermion} \tilde N_{\rm boson} N_{\rm fermion}
$, where $M$ is at a GUT scale, and $\phi_\alpha, \phi_\beta$ and $\phi_\gamma$ are the MSSM chiral fields. For this interaction, a quintic superpotential can be written as $\sim \frac{1}{M^2}N^2 W_{\rm MSSM} $ where $W_{\rm MSSM}$ is the dimension 3 superpotential.

%%%%%%%%%%%%%%%%%%%%%%%%%%%%%%%%%%%%%%%%%%%%%%%%%%%%%%%%%
%%%%%%%%%%%%%%%%%%%%%%%%%%%%%%%%%%%%%%%%%%%%%%%%%%%%%%%%%%%%%

A simpler form compared to (\ref{eq:effderInt}) can be taken as
\begin{equation}
\frac{A }{M^2_{IJ}} \tilde N^*\overline{e_J^c}
N  \tilde e_I^c  +{\rm h.c.}\label{eq:IntAterm}
\end{equation}
If $\tilde N$ of Eq. (\ref{eq:effderInt}) develops a vacuum expectation value(VEV) $V$, then we obtain the form (\ref{eq:IntAterm}) with $A=m_J$ and $\langle \tilde N^*\rangle=V^*$.  In fact, there exists a diagram, leading to Eq. (\ref{eq:IntAterm}) with the $A$-term insertion, as shown in Fig. \ref{fig:dim6HKK}(b). The importance of this diagram depends on how SUSY is broken, and it is probably more important than Fig. \ref{fig:dim6HKK}(a)  in the gravity mediation scenario. We find two more diagrams comparable to but not exceeding Fig. \ref{fig:dim6HKK}(b): one type with $m_N/A$ times Fig. \ref{fig:dim6HKK}(b) and the other with the insertion of the $B$-term $Bm_E\tilde E^c\tilde E$. In this paper, we study the form (\ref{eq:IntAterm}) for a concrete discussion.

Let us introduce $X_{1,2}$ to break the PQ symmetry \cite{Kim84} by their VEVs. Axino is predominantly $\tilde X$, a combination of $\tilde X_1$ and $\tilde X_2$. The U(1)$_\Gamma-$SU(3)$_{\rm color}-$SU(3)$_{\rm color}$ anomaly($\Gamma$CC anomaly) is present either by the heavy quark(s)  and/or the SM quark(s). The quantum numbers (the R-parity, the hypercharge $Y$, the PQ charge $\Gamma$, and the matter parity $P$) of the needed fields are listed in Table \ref{tab:chargeslep}, including the Higgs doublet pair $\phi_u$ and $\phi_d$. The coupling $f'NnX_1$ gives the first term of Eq. (\ref{eq:axinodecay}) with $M'=m_n/f'^2$. The lightest among $E_\alpha$ is called $E$ which has a huge mass splitting from the other $E_\alpha$s, but here $E$ is still treated as super-heavy toward a DDM scenario, unlike in Ref. \cite{HuhKK08}. $\Gamma$s could be assigned such that families are not distinguishable, nevertheless in Table \ref{tab:chargeslep} we devised a scheme toward a much smaller electron mass compared to the muon and tau masses.

%%%%%%%%%%%%%%%%%%%%%%%%%%%%%%%%%%%%%%%%%%%%%%%%%%%%%%%%%%%%%%%%%%%%%%%%%%%
\begin{table}[!]
\begin{center}
\begin{tabular}{c||c|cccccc||cccc}
\hline\hline
&~$N$~& $n$& $X_1$ &$ X_2$&$\Sigma$&~$E$  &~$E^c$ &$\ell_{I}$&$e_{I}^c$ &$\phi_u$&$\phi_d$
\\[0.1em]
\hline &&&&&&&&&&&\\[-1.3em]
$R$~ & $+$&$+$& $+$& $+$&$+$&$-$& $-$ &$-$&$-$&$+$&$+$\\[0.3em]
$Y$~ & $0$&$0$ &$0$ &$0$&$0$ &$-1$ &$+1$ &$-\frac12$&$+1$&$+\frac12$&$-\frac12$
\\ [0.2em]
\hline
$\Gamma$~ & $+1$&$0$& $-1$ &$+1$&$+2$ &$0$&$0$ &$0,-1,-1$~~&$1,1,1$~ &$+2$&$0$
\\ [0.3em]
$P$~ & $-$&$-$& $+$  &$+$&$+$ &$-$&$-$ &~~$+,+,+$~~&$+,+,+$~ &$+$&$+$
\\ [0.2em]
\hline
\end{tabular}
\end{center}
\caption{Color singlet chiral fields and their quantum numbers. } \label{tab:chargeslep}
\end{table}

Toward the $\mu$ term generation {\it \`a la} Giudice and Masiero \cite{Giudice88}, we also listed $\Sigma$ in Table \ref{tab:chargeslep} for the K\"ahler potential term $\Sigma^*\phi_u\phi_d/M_P$. The following renormalizable superpotential is consistent with the symmetries of Table \ref{tab:chargeslep}, for $I=e,\mu,\tau$ and $\alpha=1,2,3$,
\dis{
&W=f_{I\alpha}Ne_IE_\alpha^c+m_E^\alpha E_\alpha E^c_\alpha+\frac12 m_nn^2 +\lambda X_1^2\Sigma\label{eq:Wfull}
}
where the first term is the coupling considered in \cite{HuhKK08}.

\begin{figure}[!]
\begin{center}
\resizebox{1\columnwidth}{!}
{\includegraphics{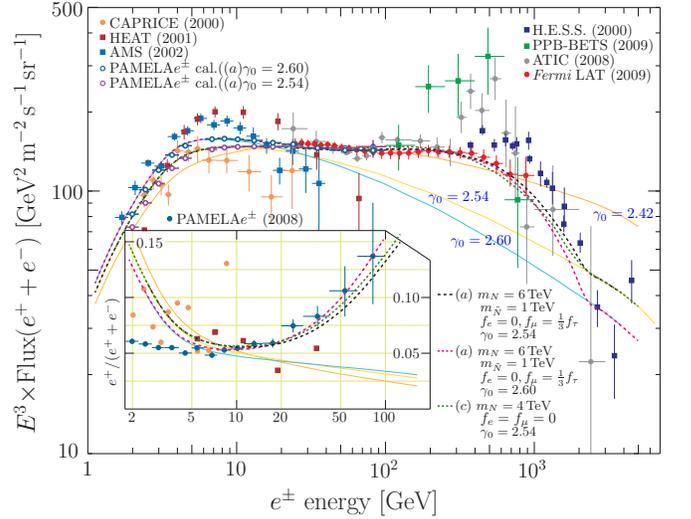}}
\end{center}
\caption{A few fits of DDM masses to CR $e^\pm$ with $f_\ell\equiv f_{\ell E}$. The slepton masses are those given at the benchmark points of SPS1a \cite{SPS1}. The skyblue (goldenrod) line is the $e^\pm$ background for Case ($a$) with $\gamma_0 = 2.60 (2.54)$ accounting the galactic propagation through the GALPROP \cite{GALPROP}. The dandelion line is for $\gamma_0 = 2.42$.  Propagation of excess $e^\pm$ from the DDM decay is calculated by the GALPROP code. The solar modulation potential relevant near the GeV region is $\Phi=500$ MV. The data are from the CAPRICE(peach dots) \cite{CAPRICE00}, the AMS(navyblue squares) \cite{AMS01}, the HEAT(maroon squares) \cite{HEAT01}, the PAMELA$e^\pm$ ratio(midnightblue dots) in the inset \cite{PAMELAe}, the calculated  PAMELA$e^\pm$ with $\gamma_0=2.54,2.60$(purple, midnightblue circles), H.E.S.S.(blue squares) \cite{HESS08}, PPB-BETS(green squares) \cite{PPBBETS2008, BETS01},  ATIC(gray dots) \cite{ATICnature}, and the {\it Fermi} LAT(red dots) \cite{Fermi1}. }
\label{fig:eespectra}
\end{figure}

\vskip 0.3cm
%%%%%%%%%%%%%%%%%%%%%%%%%%%%%%%%%%%%%%%%%%%%%%%%%%%%%%%%%
{\it Lifetime of {\rm $N$}~}---
The TeV scale sector contains fields $\tilde N, N$, saxion $s$ and axino $\tilde a$, whose mass hierarchy is assumed to be
\begin{equation}
m_\chi \ll m_N , m_{\tilde N}< m_s, m_{3/2}, m_{\tilde a}, \label{eq:masshier}
\end{equation}
where $m_{3/2}$ is the gravitino mass,
and we use the following VEV of $\tilde N$ and the suppression mass,
 \begin{equation}
\langle\tilde N\rangle\simeq V, \quad M^2_{IJ}=\left|\frac{m_E^2}{f_If^*_J}\right|.
\end{equation}
We distinguish the three cases for the interaction (\ref{eq:IntAterm}) according to the masses of the $N$ supermultiplet and the VEV of $\tilde N$. The discussion on how  $\tilde N$ develops a VEV or not is outside the scope of this paper, and below we will simply choose the cases of $V=0$ or $V\ne 0$ toward a phenomenological study. In Fig. \ref{fig:eespectra}, we show some parameters fitted to the observed flux.
%%%%%%%%%%%%%%%%%%%%%%%%%%%%%%%%%%%%%%%%%%%%%%%%%%%%%%%%%%%%%%%%%%%%%%%%%%%%

\vskip 0.2cm
\underline{Case ({\it a}): $m_N> m_{\tilde N}, V=0$:}
The matter parity $P$ is unbroken and $\tilde N$ is a stable particle. If gravitino is lighter than the $N$ multiplet, $N$ can decay to gravitino with the amplitude suppressed by one power of the Planck mass. So, we forbid this by the hierarchy, (\ref{eq:masshier}). By interaction (\ref{eq:IntAterm}), $N$ decays to ${\tilde N}^*+ {e_I^c}+ \tilde e_J^{c*} $, and to their charge conjugated states. So, we estimate the decay width of $N$ as
\begin{equation}
\begin{array}{l}
\Gamma=\frac{A^2 m_N^3}{2^6\pi^3 M^4}\int_{\sqrt{y}}^{\xi_{\rm max}}d\xi\,
\frac{(1-x+y-2\xi)^2(1-\xi)\sqrt{\xi^2-y}}{(1+y-2\xi)^2},\\[0.7em]
 \quad \quad \quad \quad \xi_{\rm max}=\frac12(1-x+y)
\end{array}\label{eq:Casea}
\end{equation}
where $x=m^2_{\tilde N}/m^2_N, y=m^2_{\tilde \tau}/m^2_N$ and we take a real $A$ which is of order $m_{3/2}$. In the limit of  $x\to 0$ and $ y\to 0$, we obtain $\Gamma= A^2 m_N^3/768\pi^3 M^4$.
For Eq. (\ref{eq:Casea}) to give an order of $ 10^{26}(m_N/m_\chi)$ s  for
$m_\chi=100$ GeV and $A=1~\tev$, we need a relation, $M\sim 2.8\times 10^{15}(m_N/{\tev})~\gev$.

%%%%%%%%%%%%%%%%%%%%%%%%%%%%%%%%%%%%%%%%%%%%%%%%%%%%%%%%%

\vskip 0.2cm
%%%%%%%%%%%%%%%%%%%%%%%%%%%%%%%%%%%%%%%%%%%%%%%%%%%%%%%%%%%%%%%%%%%%%%%%%%%
\underline{Case ({\it b}) $m_{\tilde N}> m_N, V=0$:}
The matter parity $P$ is unbroken and $N$ is a stable particle. The $\tilde N$ decay to gravitino is forbidden by the hierarchy, (\ref{eq:masshier}). By interaction (\ref{eq:IntAterm}), $\tilde N$ decays to ${N}+{e_I^c}+ \tilde e_J^{c*}$, and to their charge conjugated states. So, we estimate the decay width of $\tilde N$ as
\begin{equation}
\begin{array}{l}
\Gamma=\frac{A^2 m_{\tilde N}^3}{2^6\pi^3 M^4}\int_{\sqrt{\tilde{x}}}^{\eta_{\rm max}}d\eta \frac{(1+\tilde x-\tilde y-2\eta)^2(\eta-\tilde x)\sqrt{\eta^2-\tilde x}}{(1+\tilde x-2\eta)^2},\\[0.7em]
 \quad \quad \quad \quad \eta_{\rm max}=\frac12(1+\tilde x-\tilde y)
\end{array}\label{eq:Caseb}
\end{equation}
where $\tilde x=m^2_N/m^2_{\tilde N}$ and $ \tilde y=m^2_{\tilde \tau}/m^2_{\tilde N}$. In the limit of $\tilde x=\tilde y=0$, we have $\Gamma=(A^2m^3_{\tilde N}/1536\pi^3 M^4)$. For Eq. (\ref{eq:Caseb}) to give an order of $2\times 10^{26}(m_{\tilde N}/m_\chi)$ s (another factor 2 for both $\tilde N$ and $\tilde N^*$ decays), we need  $M\sim  2.8\times 10^{15}(m_{\tilde N}/\tev)~\gev$ for  $\tilde x\to 0,  \tilde y\to 0$ and $m_\chi=100$ GeV and $A=1~\tev$.

\vskip 0.2cm
%%%%%%%%%%%%%%%%%%%%%%%%%%%%%%%%%%%%%%%%%%%%%%%%%%%%%%%%%%%%%%%%%%%%%%%%%%%%%%
 \underline{Case ({\it c}) $m_{\tilde N}> m_N, V\ne 0$:}
This is the simplest case. The matter parity $P$ is broken by the VEV $V$. Since we introduced only one global symmetry U(1)$_\Gamma$, the EW scale VEV of the PQ charge carrying field $\tilde N$ does not lead to any other Goldstone boson in addition to the one already introduced at the scale $F_a$.
The lightest $P$ odd particle $N$ decays to the MSSM particles. DDM is the fermion $N$  which can decay by the interaction (\ref{eq:IntAterm}): $N^c(=N)\to {e_I^c}+ \tilde e_J^{c*}$, and to their charge conjugated states. So, we estimate the decay width as
\dis{
\Gamma(N\to {e_I^c} \tilde e_J^{c*}, \overline{e}_I^c \tilde e_J^{c})\simeq \frac{V^2A^2 m_N}{16\pi M^4}\left(1-\frac{m_{\tilde e_J}^2}{m_{N}^2}\right)^2,\label{eq:Casec}
}
which becomes $3\times 10^{-23}{\rm s}^{-1}
({10^{15} \gev}/{ M})^4 ({ V}/100\gev)^2$\\ $\cdot ({ A} /{10\tev})^2 ({m_N}/{\tev})$,
where we neglected the slepton mass. To give an order of $10^{26}(m_N/m_\chi)$ s, we need a relation
$M\sim 7.4\times 10^{15}~\gev(m_NV/\tev^2)^{1/2}$ for  $m_\chi=100$ GeV.

%\vskip 0.3cm
%%%%%%%%%%%%%%%%%%%%%%%%%%%%%%%%%%%%%%%%%%%%%%%%%%%%%%%%%%%%%
%%%%%%%%%%%%%%%%%%%%%%%%%%%%%%%%%%%%%%%%%%%%%%%%%%%%%%%%%%%%%%
%{\it Fitting to CR electrons and positrons~}---
\section{Fitting to CR electrons and positrons}
In Fig. \ref{fig:eespectra}, we present the best fit DDM masses at the SPS1a benchmark point \cite{SPS1} for Cases ($a,c$). Cases ($a$) and ($b$) with the exchange $N\leftrightarrow\tilde N$ are almost indistinguishable. The {\it Fermi} LAT data may be fitted by a different injection spectrum $\gamma_0=2.42$ (the dandelion line), but then the PAMELA data is far above this dandelion curve as shown in the inset \cite{Profumo09}. The PAMELA data does not give an independent flux for $(e^++e^-)$, and hence we calculate the total flux from the ratio, $r=e^+/(e^++e^-)$, using the calculated background estimates of the goldenrod and skyblue  curves for Case ($a$) \cite{GALPROP}. The midnightblue dots
in the inset go to the purple($\gamma_0=2.54$) and midnightblue($\gamma_0=2.60$) circles for $e^++e^-$. The production rates of $e^\pm$ from the DDM decay are calculated using the  isothermal profile. PYTHIA is used to obtain the $e^\pm$ spectrum from the decay of DDM. The galactic propagation of these CR $e^\pm$ (from the DDM decay and the local sources) to Earth is estimated using the CR propagation package GALPROP \cite{GALPROP}, partially modifying it.
The parameters of the fitted curves are as shown in the figure. For example, Case ($a$) with the magenta dashed line is an excellent fit with $m_N=6\,$TeV, $m_{\tilde N}=1\,$TeV at a benchmark point of the SPS1a  \cite{SPS1} with the couplings $f_{eE}=0$ and $f_{\mu E}=\frac13f_{\tau E}$. From the figure, we notice that the fit is a combination of $\gamma_0$, the DDM mass and the couplings.

%%%%%%%%%%%%%%%%%%%%%%%%%%%%%%%%%%%%%%%%%%
%%%%%%%%%%%%%%%%%%%%%%%%%%%%%%%%%%%%%%%%

%{\it Conclusion~}---
\section{Conclusion}
We introduced just one more chiral multiplet $N$ beyond the MSSM particles at the next higher mass level of TeV, which allows the $N$ lifetime in the $10^{27}$ s range by dimension 6 operators.  We have successfully fitted both the PAMELA and {\it Fermi} LAT data with $e^\pm$ produced by the decay of $N$.  The heavy axino, decaying to both $N$ and $\chi$ below the $\chi$ decoupling temperature, enables us to estimate the relative abundances of $N$ and $\chi$. An interesting aspect of this axino decay scenario is that the suppression mass scales considered in Eqs. (\ref{eq:axinodecay},\ref{eq:Casea},\ref{eq:Caseb},\ref{eq:Casec}) fall in the general GUT scale \MG.
%\end{widetext}

\acknowledgments
%\noindent {\bf Acknowledgments:}
{We thank Bumseok Kyae for helpful discussions.
This work is supported by the Korea Research Foundation, Grant No. KRF-2005-084-C00001.}

%%%%%%%%%%%%%%%%%%%%%%%%%%%%%%%%%%%%%%%%%%%%%%%%%%%%%%%%%

\end{document}